\documentstyle[preprint,aps,eqsecnum,amssymb,graphicx]{revtex}
\tightenlines
\begin{document}
\draft
\title{Nuclear shadowing at low photon energies}

\author{T. Falter, S. Leupold and U. Mosel}

\address{Institut f\"ur Theoretische Physik\\ Justus-Liebig-Universit\"at
Giessen\\ D-35392 Giessen, Germany}

\date{February 26, 2001}

\maketitle

\begin{abstract}
We calculate the shadowing effect in nuclear photoabsorption at low photon 
energies (1-3~GeV) within a multiple scattering approach. We avoid some 
of the high energy approximations that are usually made in simple Glauber 
theory like the narrow width and the eikonal approximation. We find that the 
main contribution to nuclear shadowing at low energies stems from $\rho^0$ 
mesons with masses well below their pole mass. We also show that the 
possibility of scattering in 
non forward directions allows for a new contribution to shadowing at low 
energies: the production of neutral pions as intermediate hadronic states 
enhances the shadowing effect in the onset region. For light nuclei and small 
photon energies they give rise to about 30\% of the total shadowing effect.
\end{abstract}
\pacs{PACS numbers: 25.20.Dc, 24.10.Ht, 12.40.Vv}


\section{Introduction} \label{sec:intro}

At high energies the total cross section for the reaction of a hadronic 
projectile with a nucleus of mass number $A$ is usually smaller 
than $A$ times the nucleonic cross section. This phenomenon, called the 
shadowing effect, can be understood if the mean free path of the 
projectile inside the nucleus is smaller than the nuclear dimension. In that 
case the nucleons on the front side of the nucleus shadow the downstream 
nucleons which therefore do not contribute to the total nuclear cross section. 
Since the nucleonic cross section of photons is smaller than typical hadronic 
ones by a factor $\alpha_{em}$ in order of magnitude it is at first sight 
astonishing
that photonuclear reactions are shadowed at high energies. A qualitative 
explanation for the shadowing of high energy photons is given by their 
property to fluctuate into a hadronic state with the quantum numbers of the
photon. This happens with a probability of order $\alpha_{em}$. 
The distance that such a hadronic fluctuation travels, the so 
called coherence length, can be estimated from the uncertainty principle: 
\begin{equation}
  \label{eq:coherence}
  l_h\approx\left|k-\sqrt{\nu^2-m_h^2}\right|^{-1}
\end{equation}
where $k$ denotes the momentum of the photon, $\nu$ its energy and $m_h$
the mass of the hadronic fluctuation ($k=\nu$ for real photons). If the 
coherence length becomes larger 
than the mean free path of the hadronic fluctuation inside the nucleus its 
total nuclear cross section and therefore also that of the photon is 
shadowed.\\

\begin{figure} 
\begin{center}
\includegraphics[width=10cm]{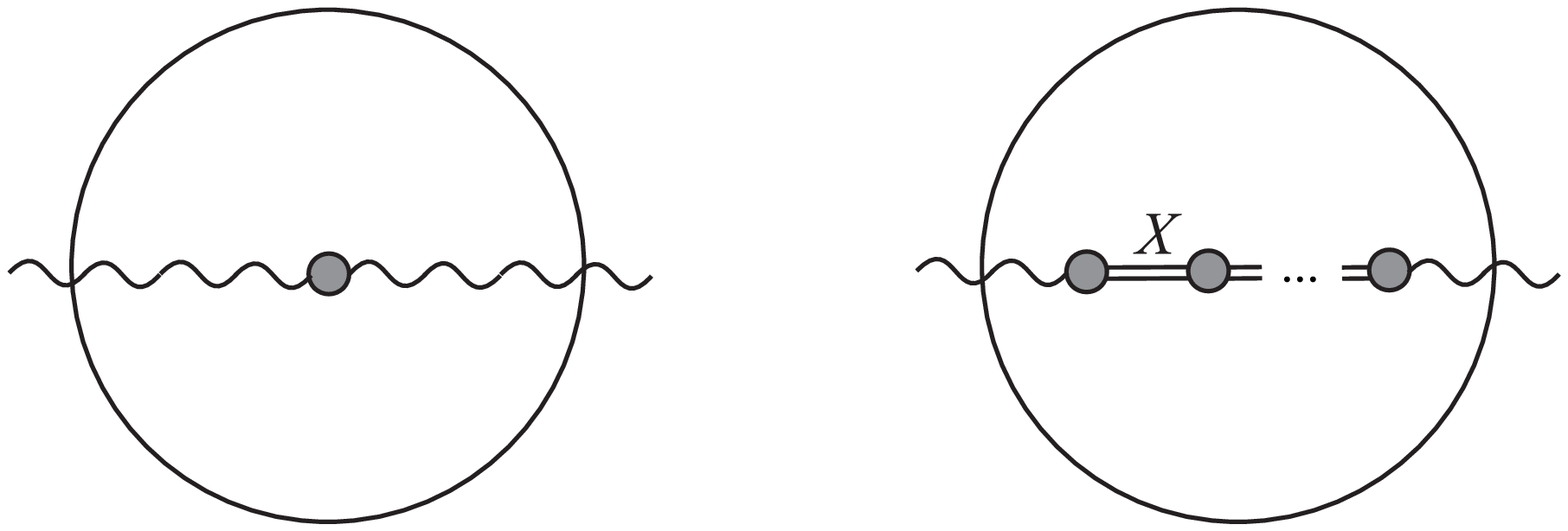}
\end{center}
\caption{The two amplitudes that contribute in order $\alpha_{em}$ to the 
nuclear Compton forward scattering amplitude in the Glauber model. The left 
amplitude alone would lead to an unshadowed cross section. The right one gives
rise to nuclear shadowing.}
\label{fig:glauber}
\end{figure}

Quantitatively, the shadowing of nuclear photoabsorption can be
understood within the simple Glauber model \cite{Gl59,Gl70,Yen71} 
by relating the nuclear photoabsorption cross section via the optical theorem 
to the nuclear Compton forward scattering amplitude. In order $\alpha_{em}$ 
one gets two contributions to this amplitude (see Fig~\ref{fig:glauber}). The 
first one stems from forward scattering of the photon from a single nucleon
inside the nucleus. Taking only this contribution into account leads to an 
unshadowed nuclear cross section. Its interference with the second amplitude 
in order $\alpha_{em}$ causes the shadowing effect. Assuming that at high 
energies all scattering events are dominated in the forward direction 
(eikonal approximation) this second  amplitude takes the following form: The 
incoming photon produces an intermediate hadronic state $X$ at one nucleon 
without exciting the nucleus. Within the eikonal approximation this process is
presumed to happen in the forward direction so that the hadronic state must 
have the quantum numbers of the photon. One therefore usually expects that 
these states are dominated by vector mesons. The intermediate hadronic state 
then scatters at fixed impact parameter
through the nucleus and finally at some nucleon into the outgoing photon. One 
assumes that during the whole process the nucleus stays in its ground state 
and neglects processes with intermediate excitation of the nucleus. This so 
called multiple scattering approximation corresponds roughly to neglecting two
body and higher nuclear correlations and is expected to be reasonable
for high energies (see e.g.~\cite{Ker59,Fes59}).\\

Within the Glauber approach one can quantitatively understand nuclear 
shadowing of high energy photons (see e.g. \cite{Bau78,Don78} and references 
therein), but recent photoabsorption data \cite{Bia96,Muc98} on C, Al, Cu, Sn 
and Pb in the energy range from 1 to 2.6~GeV display an early onset of the 
shadowing effect which some of the newer models \cite{Pil95,Bof96,Eng97} 
cannot explain. In \cite{Fal00} we showed that one can understand the
data within the simple Glauber model by taking the negative real part of the 
$\rho^0 N$ scattering amplitude into account. However, some of the assumptions
that are made in this simple model, like the validity of the eikonal 
approximation and the neglect of the finite width of the $\rho^0$ meson become 
questionable at low energies. In the present work we therefore use a multiple 
scattering approach~\cite{Gri70,Ber72,Wei76} that we consider as more reliable
in the shadowing onset region. Within this approach one can account for
the vacuum self-energy of the $\rho^0$ and include scattering processes in 
nonforward direction. We find that the latter allow for a new contribution to 
nuclear shadowing in photoabsorption at low energies, namely the production of 
$\pi^0$ as intermediate hadronic states. Since neutral pions cannot be 
produced in the forward direction without exciting the nucleus they 
do not contribute in calculations based on the eikonal approximation.\\

In Sec.~\ref{sec:model} we give a brief description of the multiple scattering
formalism as a method for calculating the nuclear Compton forward scattering 
amplitude from free scattering amplitudes. The calculated ratio 
$\sigma_{\gamma A}/A \sigma_{\gamma N}$ that we obtain 
within this model is presented in Sec.~\ref{sec:results}.
If one only considers vector mesons as intermediate hadronic states and 
accounts for the finite width of the $\rho^0$ meson the 
results are already in good agreement with the experimental data.
In Sec.~\ref{sec:pion} we show that the shadowing effect is enhanced at photon 
energies around 1~GeV by taking the contribution of the $\pi^0$ as an 
intermediate state into account. We summarize our results in 
Sec.~\ref{sec:summary}.\\


\section{Glauber-Gribov multiple scattering formalism} \label{sec:model}
To calculate the photoabsorption cross section of a nucleus with mass number 
$A$ we make use of the optical theorem. Assuming that the projectile 
scatters from individual nucleons inside the nucleus we can express the 
nuclear photoabsorption cross section in terms of multiple scattering 
amplitudes 
${\mathcal{A}}^{(n)}$:
\begin{equation}
  \label{eq:optical}
  \sigma_{\gamma A}=\frac{1}{2m_Nk}\textrm{Im}\sum_{n=1}^A{\mathcal{A}}^{(n)},
\end{equation}
where $k$ denotes the momentum of the photon and $n$ the number of nucleons 
that participate in each multiple scattering process. Our goal is to calculate 
the amplitudes ${\mathcal{A}}^{(n)}$ from free nucleonic scattering amplitudes
${\mathcal{M}}$:
\begin{equation}
    \langle \vec p';\vec k'|iT|\vec k;\vec p\rangle=(2\pi)^4\delta^{(4)}(p+k-(p'+k'))\cdot i{\mathcal{M}}_\gamma(\vec k,\vec p\rightarrow \vec k',\vec p')
\end{equation}
where $\vec p$ ($\vec p'$) and $\vec k$ ($\vec k'$) denote the momenta of the
incoming (outgoing) nucleon and projectile respectively.
We therefore have to express the bound nucleon states in terms of free nucleon
states $|\vec p\rangle$ of momentum $\vec p$. The bound nucleons are treated 
nonrelativistically and we assume that they are described by nonrelativistic 
one-particle wave functions
\begin{equation}
  \label{eq:wavefct}
    \psi_\alpha(x)=\psi_\alpha(\vec x)e^{-iE_\alpha t}.
\end{equation}
The probability amplitude for finding the momentum $\vec p$ in the bound 
nucleon state $|\psi_\alpha\rangle$ is given by the Fourier transform 
$\tilde{\psi}_\alpha(\vec p)$ of (\ref{eq:wavefct}) with the normalisation
\begin{equation}
  \int\frac{d^3p}{(2\pi)^3}|\tilde{\psi}_\alpha(\vec p)|^2=1.
\end{equation}
Therefore we can write for the bound nucleon states
\begin{eqnarray}
  |\psi_\alpha\rangle&=&\int\frac{d^3p}{(2\pi)^3}\tilde{\psi}_\alpha(\vec p)\frac{1}{\sqrt{2E_{\vec p}}}|\vec p\rangle\nonumber\\
  &\approx&\int\frac{d^3p}{(2\pi)^3}\tilde{\psi}_\alpha(\vec p)\frac{1}{\sqrt{2m_N}}|\vec p\rangle.
\end{eqnarray}
The factor of $\sqrt{2E_{\vec p}}$ converts the relativistic normalization of 
$|\vec p\rangle$ 
\begin{equation}
  \langle\vec p|\vec p'\rangle=2E_{\vec p}(2\pi)^3\delta^{(3)}(\vec p-\vec p')
\end{equation}
to the conventional normalization $\langle\psi_\alpha|\psi_\alpha\rangle=1$.
\\

\begin{figure} 
\begin{center}
\includegraphics[width=6cm]{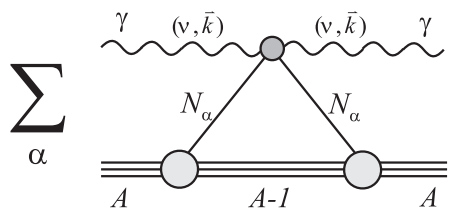}
\end{center}
\caption{First contribution to the photon-nucleus forward scattering amplitude
in order $\alpha_{em}$: Forward scattering of the incoming photon from a bound
 nucleon $N_\alpha$. This amplitude corresponds to the left amplitude in 
Fig.~\ref{fig:glauber} and leads to an unshadowed cross section.}
\label{fig:single-scattering}
\end{figure}

The amplitude ${\mathcal{A}}^{(1)}$ for forward scattering of a photon
with momentum $\vec k$ and energy $\nu$($=|\vec k|$ for real photons) from a 
single bound nucleon (see Fig.~\ref{fig:single-scattering}) then takes the form
\begin{equation}
  i{\mathcal{A}}^{(1)}=\sum_{\alpha=1}^A\int\frac{d^3p}{(2\pi)^3}\tilde{\psi}_\alpha^*(\vec p)i{\mathcal{M}}_\gamma(\vec k,\vec p\rightarrow\vec k,\vec p)\tilde{\psi}_\alpha(\vec p)
\end{equation}
where ${\mathcal{M}}_\gamma$ is the free invariant photon nucleon scattering 
amplitude.
We use the local approximation and neglect the dependence of the amplitude 
${\mathcal{M}}_\gamma$ on the momentum 
$\vec p$ of the incoming nucleon, i.e. we make the replacement
 \begin{equation}
   {\mathcal{M}}_\gamma (\vec k,\vec p\rightarrow\vec k,\vec p)\rightarrow{\mathcal{M}}_\gamma (\vec k,\vec p_0\rightarrow\vec k,\vec p_0).
\end{equation}
Since the momentum of the photon $\vec k$ is much larger than the Fermi 
momentum of the bound nucleons, we set $\vec p_0\approx\vec 0$ and the 
amplitude ${\mathcal{A}}^{(1)}$ takes the simple form
\begin{eqnarray}
  i{\mathcal{A}}^{(1)}&=&i{\mathcal{M}}_\gamma(\vec k,\vec p_0\rightarrow\vec k,\vec p_0)\sum_{\alpha=1}^A\int\frac{d^3p}{(2\pi)^3}|\tilde{\psi}_\alpha(\vec p)|^2\\
  &=&Ai{\mathcal{M}}_\gamma(\vec k,\vec 0\rightarrow\vec k,\vec 0)
\end{eqnarray}
which leads to the unshadowed cross section $A\sigma_{\gamma N}$ when used in 
Eq.~(\ref{eq:optical}).\\

\begin{figure} 
\begin{center}
\includegraphics[width=9cm]{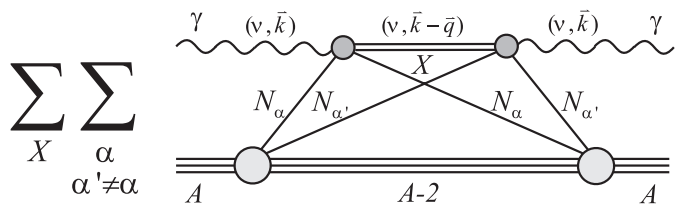}
\end{center}
\caption{Second contribution to the photon-nucleus forward scattering amplitude
in order $\alpha_{em}$: The photon produces a hadron $X$ at nucleon $N_\alpha$.
This hadron propagates freely to a second nucleon $N_{\alpha '}$ where it 
scatters into the outgoing photon. We assume that the nucleus stays in its 
ground state during the whole scattering process, i.g. there is no energy 
transfered to the bound nucleons.}
\label{fig:double-scattering}
\end{figure}

The double scattering amplitude ${\mathcal{A}}^{(2)}$ is shown in 
Fig.~\ref{fig:double-scattering} and corresponds to the process where the 
incoming photon produces a hadronic state $X$ which propagates freely without 
further scattering to a second nucleon and there scatters into the outgoing 
photon. We assume that the nucleus stays in its ground state, i.e. we 
neglect nondiagonal contributions with an excited nucleus between the two 
scattering events (multiple scattering approximation). Hence there is no 
energy transfer to the two nucleons. If $\vec q$ denotes the momentum transfer
to the first nucleon the fact that we are dealing with the nuclear forward 
scattering amplitude fixes the momentum transfer to the second nucleon to 
$-\vec q$. Using again the local approximation and introducing the nuclear 
formfactor
\begin{equation}
  \label{eq:formfac}
  F(\vec q)=\int d^3xe^{i\vec q\cdot\vec x}|\psi_\alpha(\vec x)|^2=\frac{1}{A}\int d^3xe^{i\vec q\cdot\vec x}n(\vec x)
\end{equation}
where $n(\vec x)$ denotes the nucleon number density, the double scattering 
amplitude takes the simple form
\begin{eqnarray}
  \label{eq:double}
  i{\mathcal{A}}^{(2)}&=&A(A-1)\int\frac{d^3q}{2m_N(2\pi)^3}\nonumber\\
  & &\times\sum_Xi{\mathcal{M}}_{\gamma X}(\vec q)F(\vec q)\frac{i}{\nu^2-(\vec k-\vec q)^2-\tilde{m}_X^2-\Pi_X(\nu^2-(\vec k-\vec q)^2)}i{\mathcal{M}}_{X\gamma}(\vec q)F(-\vec q).
\end{eqnarray}
Here $\tilde{m}_X$ and $\Pi_X$ denote the bare mass and the vacuum self energy of
the intermediate state $X$ and we used the following abbreviation for the invariant production amplitudes:
\begin{eqnarray}
  i{\mathcal{M}}_{\gamma X}(\vec q)&=&i{\mathcal{M}}_{\gamma X}(\vec k,\vec p_0\rightarrow\vec k-\vec q,\vec p_0+\vec q)\\
  i{\mathcal{M}}_{X\gamma}(\vec q)&=&i{\mathcal{M}}_{X\gamma}(\vec k-\vec q,\vec p_0\rightarrow\vec k,\vec p_0-\vec q).
\end{eqnarray}
The nuclear formfactor $F(q)$ suppresses large values of $q=|\vec q|$ and 
therefore favors light intermediate states. Though it is in principle possible
to produce slow, massive $\rho^0$ mesons, their contributions are suppressed by
the nuclear formfactor. The important contributions stem from light $\rho^0$ 
mesons having the energy of the photon ($>1$~GeV) so that the effect of Fermi 
motion becomes negligible. We therefore set again $\vec p_0\approx\vec 0$.
In Sec.~\ref{sec:results} we will show that double scattering gives the 
leading contribution to nuclear shadowing.\\

\begin{figure} 
\begin{center}
\includegraphics[width=14cm]{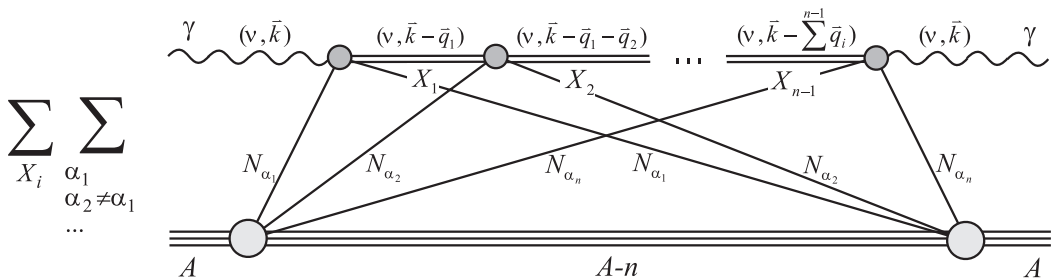}
\end{center}
\caption{General form of ${\mathcal{A}}^{(n)}$ for $n\geq 1$.}
\label{fig:nfold-scattering}
\end{figure}

The form of the $n$-fold scattering amplitude ${\mathcal{A}}^{(n)}$ can be 
read off from Fig.~\ref{fig:nfold-scattering}. Using the same approximations 
as before one gets
\begin{equation}
  i{\mathcal{A}}^{(n)}=\frac{A!}{(A-n)!}\prod_{i=1}^{n-1}\left[\int\frac{d^3q_i}{2m_N(2\pi)^3}\right]F(\vec q_1)...F(\vec q_n)i{\mathcal V}^{(n)}(\{\vec q_i\})
\end{equation}
with
\begin{eqnarray}
  \label{eq:vn}
  i{\mathcal V}^{(n)}(\{\vec q_i\})&=&\sum_{X_i}i{\mathcal{M}}_{\gamma X_1}(\vec q_1)\frac{i}{\nu^2-(\vec k-\vec q_1)^2-\tilde{m}_{X_1}^2-\Pi_{X_1}(\nu^2-(\vec k-\vec q_1)^2)}i{\mathcal{M}}_{X_1X_2}(\vec q_2)\nonumber\\
    & &\qquad\times\frac{i}{\nu^2-(\vec k-(\vec q_1+\vec q_2))^2-\tilde{m}_{X_2}^2-\Pi_{X_2}(\nu^2-(\vec k-(\vec q_1+\vec q_2))^2)}\nonumber\\
    & &\qquad\qquad\times...\quad i{\mathcal{M}}_{X_{n-1}\gamma}(\vec q_n)
  \end{eqnarray}
and
\begin{eqnarray}
  \vec q_n&=&-\sum_{i=1}^{n-1}\vec q_i\\
  {\mathcal{M}}_{\alpha\beta}(\vec q_i)&=&{\mathcal{M}}_{\alpha\beta}(\vec k-\sum_{j=1}^{i-1}\vec q_j,\vec 0\rightarrow\vec k-\sum_{j=1}^{i}\vec q_j,\vec q_i).
\end{eqnarray}
Using the eikonal approximation within this approach corresponds to the 
neglect of any dependence of ${\mathcal V}^{(n)}$ on the momentum transfer 
$\vec q_{\bot,i}$ transverse to the initial photon direction, i.e. setting
$\vec q_{\bot,i}=0$ in Eq.~(\ref{eq:vn}) (see e.g. \cite{Ber72}).
At high energies this is certainly a good approximation, so that for high
energies and in the large $A$ limit this multiple scattering series can be
summed up and reduces to the simple Glauber formula used in~\cite{Fal00}
if one neglects the widths of the vector mesons. To describe photoabsorption 
also at lower energies in the following we will relax the eikonal 
approximation by allowing nonforward scattering events.\\


\section{Beyond the eikonal approximation} \label{sec:results}
The nuclear formfactor $F(\vec q)$ suppresses large momentum transfer 
$\vec q$. Hence at very high energies the scattering processes tend 
predominantly into the forward direction, justifying the usage of the eikonal 
approximation. In the eikonal limit the intermediate hadronic states $X_i$ in 
(\ref{eq:vn}) must have the quantum numbers of the photon since they are 
produced without nuclear excitation. With decreasing energy nonforward 
scattering becomes more important. In this section we present a calculation 
that goes beyond the eikonal approximation but we still restrict the 
intermediate hadronic states $X_i$ to the light vector mesons 
$V=\rho^0,\omega,\phi$ that are the only possible intermediate states
in the eikonal limit. Note that the nuclear formfactor suppresses the 
production of more massive states in the energy regime we are 
interested in. In Sec.~\ref{sec:pion} we will see that at photon energies 
around 1~GeV it becomes also important to have a $\pi^0$ as an intermediate 
state. In the following we will neglect nondiagonal contributions where a state
$X_i$ scatters into a different state $X_j$.\\

For $\omega$ and $\phi$ mesons we use constant self energies such that:
\begin{eqnarray}
  \tilde{m}_{\omega,\phi}^2+\textrm{Re}\Pi_{\omega,\phi}&=&m_{\omega,\phi}^2\\
  \textrm{Im}\Pi_{\omega,\phi}&=&-m_{\omega,\phi}\Gamma_{\omega,\phi},
\end{eqnarray}
with the physical masses $m_\omega$ and $m_\phi$ and the widths $\Gamma_\omega$
and $\Gamma_\phi$ taken from~\cite{PDG98}. For the $\rho^0$ we use a 
momentum dependent vacuum self energy $\Pi_\rho(p^2)$~\cite{Kli96}.\\

The nuclear formfactor $F(q)$ is calculated via Eq.~(\ref{eq:formfac}) 
assuming a Woods-Saxon distribution
\begin{equation}
  \label{eq:woods-saxon}
  n(\vec r)=\frac{\rho_0}{1+\exp\left[\frac{r-R}{a}\right]}
\end{equation}
with the parameters listed in Tab.~\ref{tab:woods-saxon}.\\

We make the following ansatz for the elastic $VN$ scattering amplitudes, 
motivated by the angular distribution of $\rho^0$ in photoproduction 
data~\cite{ABBHHM68}:
\begin{equation}
  {\mathcal{M}}_V(s,t,m)=8\pi m_Nf_V(\vec 0,k_V(s,m))e^{\frac{1}{2}Bt}
\end{equation}
where $k_V$ denotes the momentum of the vector meson with invariant mass $m$ in
the rest frame of the nucleon at invariant energy $\sqrt{s}$. For $t=0$ 
this expression yields exactly the invariant forward scattering amplitude which
is related to the total $VN$ cross section via the optical theorem:
\begin{eqnarray}
  \sigma_{VN}&=&\frac{1}{2m_Nk_V}\textrm{Im}{\mathcal{M}}(s,t=0,m)\\
  &=&\frac{4\pi}{k_V}\textrm{Im}f_V(\vec 0,k_V(s,m)).
\end{eqnarray}
The slope parameter in the considered energy region can
be estimated from $\rho^0$ photoproduction data~\cite{ABBHHM68}:
\begin{equation}
  B\approx 6\text{ GeV}^{-2}.
\end{equation}
We use the $\rho^0 N$ forward scattering amplitude $f_\rho(\vec 0,k_V)$ 
from~\cite{Kon98}. In~\cite{Fal00} we already showed that the negative real 
part of $f_\rho$ enhances the shadowing effect at low energies and leads to a 
higher effective mass of the $\rho^0$ meson in nuclei. Also in the present 
calculation the real part of $f_\rho$ turns out to be important. We set
\begin{eqnarray}
  f_\omega(\vec 0,k_\omega)&=&f_\rho(\vec 0,k_\omega)\\
  f_\phi(\vec 0,k_\phi)&=&i\frac{k_\phi}{4\pi}\sigma_\phi,
\end{eqnarray}
assuming the total $\phi N$ cross section 
$\sigma_\phi=12\text{ mb}$~\cite{Bau78}. We relate the photoproduction 
amplitude of a vector meson of invariant mass $m$ to the elastic $VN$ 
scattering amplitudes via the vector meson dominance model (VDM) and use the 
formfactor
\begin{equation}
  \label{eq:pennerfac}
  {\mathcal{F}}(m^2,m_V^2)=\frac{\Lambda^4}{\Lambda^4+\left(m^2-m_V^2\right)^2}
\end{equation}
with a cutoff parameter $\Lambda=1.2$~GeV~\cite{Feu99} to describe the 
off-shell behavior of the photoproduction amplitude for a broad vector meson:
\begin{equation}
  \label{eq:vmd}
  {\mathcal{M}}_{\gamma V}(s,t,m)={\mathcal{M}}_{V\gamma}(s,t,m)=\frac{e}{g_V}{\mathcal{M}}_V(s,t,m){\mathcal{F}}(m^2,m_V^2).
\end{equation}
Here the first equality is given by detailed balance. The coupling constants
$g_V$ are taken from model~I of Ref.~\cite{Bau78}. Note that expression 
(\ref{eq:vmd}) is not based on experimental data. It is an off-shell 
extrapolation for the production
amplitude which in the on-shell case reproduces the experimental $\rho^0$ and 
$\omega$ photoproduction data~\cite{ABBHHM68}. A low value of $\Lambda$
suppresses the production of vector mesons that are far off their mass shell. 
However, the formfactor (\ref{eq:pennerfac}) is only important in the case of a
$\rho^0$ as an intermediate state since contributions to nuclear 
shadowing originating from off-shell $\omega$ and $\phi$ are already strongly 
suppressed by their propagators in (\ref{eq:double}) because of their much
smaller widths.\\

The photon nucleon cross section $\sigma_{\gamma N}$
is approximated for each nucleus with mass number $A$ and proton number $Z$ by
the isospin averaged cross section
\begin{equation}
  \label{eq:isospin}
  \sigma_{\gamma N}=\frac{Z\sigma_{\gamma p}+(A-Z)\sigma_{\gamma n}}{A},
\end{equation}
fitting the data on $\sigma_{\gamma p}$ and $\sigma_{\gamma n}$ in the 
considered energy region.\\

\begin{figure} 
\begin{center}
\includegraphics[width=12cm]{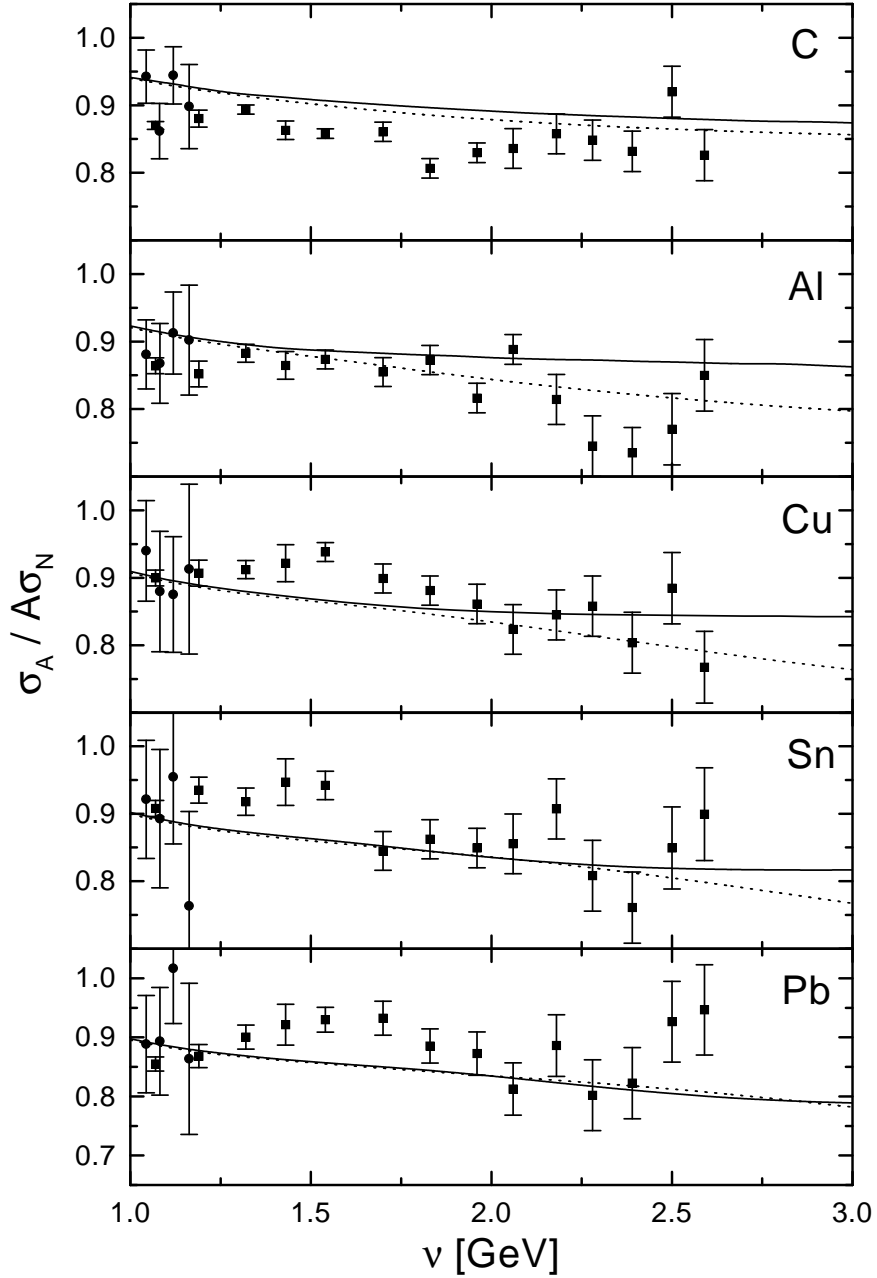}
\end{center}
\caption{Calculated ratio $\sigma_{\gamma A}/A\sigma_{\gamma N}$ as a 
function of the photon energy $\nu$. Only the light vector mesons $\rho^0$, 
$\omega$ and $\phi$ are taken into account as intermediate states. The dotted 
curve represents the calculation up to double scattering, the solid line also 
includes the corrections from triple scattering. The experimental data are 
taken from: {\small $\bullet$}~\protect\cite{Bia96}, 
{\tiny $\blacksquare$}~\protect\cite{Muc98}.}
\label{fig:noneik}
\end{figure}

In Fig.~\ref{fig:noneik} we compare the calculated ratio 
$\sigma_{\gamma A}/A\sigma_{\gamma N}$ for several nuclei with experimental 
data~\cite{Bia96,Muc98}. The dotted line represents the calculation using 
Eq.~(\ref{eq:optical}) and including terms up to double scattering ($n=2$). 
The inclusion of triple scattering ($n=3$) leads to the solid curve. The 
result including ${\mathcal{A}}^{(4)}$ lies between the two curves because 
(\ref{eq:optical}) is an alternating series as can be seen by neglecting 
the real part of the amplitudes ${\mathcal{M}}$ in (\ref{eq:vn}). Thus at low 
energies the calculation up to triple scattering is already a good 
approximation. The dominant contribution to shadowing stems from the double 
scattering term. At 
higher energies the terms with $n>3$ become important. More than 90\% of the 
shadowing is caused by the $\rho^0$ meson as an intermediate particle because 
its photoproduction amplitude and its width are larger than those of $\omega$ 
and $\phi$. At small energies light $\rho^0$ mesons ($m\ll m_\rho$), albeit 
suppressed by the propagator in (\ref{eq:double}), play the dominant role 
because the nuclear formfactor favors small invariant masses. This can be 
verified by choosing a smaller value for $\Lambda$ in (\ref{eq:pennerfac}) or
a smaller $\rho^0$ width in the propagator of (\ref{eq:double}) which in both
cases suppresses contributions from lighter $\rho^0$. In the limit of vanishing
$\rho^0$ width there will be no $\rho^0$ contributions to shadowing below the 
$\rho$ photoproduction threshold $\nu_{th}\approx 1.1$~GeV and the 
contribution above threshold will be strongly suppressed by the nuclear 
formfactor.
This can also be understood in terms of the coherence length $l_h$.
Lighter $\rho^0$ mesons have a larger coherence length as can be seen from 
Eq.~(\ref{eq:coherence}). Therefore their interactions are shadowed more than 
those of on-shell $\rho$ mesons. Note that $l_h$ is connected directly to the 
momentum transfer $|\vec q|$: $l_h=|\vec q|^{-1}$. For heavier nuclei our 
result is in good agreement with the data. The shadowing at very small photon 
energies ($\nu<1.25\text{ GeV}$) is enhanced by the inclusion of the 
$\pi^0$ as will be shown in Sec.~\ref{sec:pion}. The result within the eikonal
approximation, i.e. setting $\vec q_{\bot,i}=0$ in Eq.~(\ref{eq:vn}), leads to
a $\sim$10\% stronger shadowing in the considered energy region in case of the
vector mesons as intermediate states only. Note that additional shadowing 
stemming from intermediate $\pi^0$ is not possible in the eikonal 
approximation.\\

At the end of this section we want to discuss briefly the effect of $NN$ 
correlations. In our simple Glauber calculation~\cite{Fal00} the results were
very sensitive to two body correlations which were needed to describe the 
experimental data. The reason for this was the narrow width approximation that 
allowed only for $\rho^0$ mesons with mass $m=m_\rho=770$~MeV as intermediate 
states and led to a large momentum transfer in the production process, 
especially at low photon energies. In the present work the main contribution 
to shadowing stems from the lighter $\rho^0$ that are connected with a momentum
transfer much smaller than the characteristic momentum
$q_c=780$~MeV~\cite{Bro77} at which correlations become important. Therefore 
any correlations between the nucleons can be omitted in the present work.


\section{$\pi^0$ contribution} \label{sec:pion}
If one does not restrict the scattering events to the forward direction it 
becomes possible to produce neutral pions without exciting the nucleus. 
In this process the helicity of the nucleon at which the $\pi^0$ is produced is
not allowed to change. Decomposing the $\pi^0$ photoproduction amplitude on the
free nucleon into helicity amplitudes $H_i$~\cite{Wal69}, one can construct
the invariant photoproduction amplitude without helicity change of the nucleon:
\begin{equation}
  {\mathcal{M}}_{\gamma\pi^0}^{coh}(s,t)=\sqrt{2s}\ 8\pi(H_1(s,t)+H_4(s,t)).
\end{equation}
The helicity amplitudes are calculated using the partial wave analysis 
from Arndt et al.~\cite{Arn90}. For each nucleus we use the isospin averaged 
amplitude ${\mathcal{M}}_{\gamma N\rightarrow\pi^0N}^{coh}$ in analogy
to (\ref{eq:isospin}).
The production in forward direction is forbidden by angular momentum 
conservation. Therefore a large transverse momentum transfer is necessary to 
produce the $\pi^0$ at high energies. Since large momentum transfers are 
suppressed by the nuclear formfactor, we expect the $\pi^0$ not to contribute 
to shadowing at high energies.\\

The $\pi^0$ contribution to $\sigma_{\gamma A}$ is calculated using 
Eq.~(\ref{eq:optical}) and (\ref{eq:double}):
\begin{eqnarray}
  \label{eq:pion}
  \Delta\sigma_{\gamma A}^{(\pi^0)}&=&-\frac{A(A-1)}{2m_Nk}\textrm{Re}\int\frac{d^3q}{2m_N(2\pi)^3}\nonumber\\
  & &\quad F(\vec q)i{\mathcal{M}}_{\gamma N\rightarrow\pi^0N}^{coh}(s,t)\tilde{D}_{\pi^0}(|\vec k|^2-(\vec k-\vec q)^2)i{\mathcal{M}}_{\pi^0 N\rightarrow\gamma N}^{coh}(s,t)F(-\vec q)
\end{eqnarray}
where we have introduced the effective propagator
\begin{equation}
  \label{eq:effprop}
  \tilde{D}(p^2)=\frac{i}{p^2-m_{\pi^0}^{*2}-im_{\pi^0}^*\Gamma^*_{\pi^0}}.
\end{equation}
This propagator takes the multiple $\pi^0N$ scattering into
account. Therefore equation (\ref{eq:pion}) is not merely the double scattering
contribution with an intermediate $\pi^0$ but includes all higher order 
scattering terms within the eikonal form of the propagator (\ref{eq:effprop}).
This is equivalent to the propagation of the intermediate $\pi^0$ in an
optical potential~\cite{Fal00}. One could in principle also use the free
propagators and sum up all multiple scattering terms for the $\pi^0$ to obtain
the same result, but this is numerically more tedious. The use of the eikonal 
approximation for the intermediate $\pi^0$ scattering is justified because the
$\pi^0$ has the energy of the initial photon and therefore large momentum.
To make the same approximation for the intermediate vector mesons is not 
possible since for the energies of interest vector mesons are much slower 
because of their larger mass. The momentum dependent effective mass and the 
collisional broadening of the $\pi^0$ in the nucleus is related to the 
$\pi^0N$ forward scattering amplitude $f_{\pi^0}$~\cite{Eri88}:
\begin{eqnarray}
  m_{\pi^0}^{*2}&=&m_{\pi^0}^2-4\pi\textrm{Re}f_{\pi^0}(\vec 0,k_\pi)\rho_N\\
  m_{\pi^0}^*\Gamma^*_{\pi^0}&=&4\pi\textrm{Im}f_{\pi^0}(\vec 0,k_\pi)\rho_N.
\end{eqnarray}
The average nucleon number density
\begin{equation}
  \rho_N=\frac{1}{A}\int dr^3n(\vec r)^2
\end{equation}
is determined separately for each nucleus using Eq.~(\ref{eq:woods-saxon}) and
the parameters listed in Tab.~\ref{tab:woods-saxon}. For $\pi^0$ momenta
$k_\pi$ up to 1.65~GeV ($\sqrt{s}\leq 2\text{ GeV}$) we take the partial waves 
from the SM95 solution of Ref.~\cite{CNS00} to construct the $\pi^0N$ forward 
scattering amplitude $f_{\pi^0}$. For higher $\pi^0$ momenta we use the KA84 
solution~\cite{CNS00}. Note that $m_{\pi^0}^{*2}$ as well as 
$m_{\pi^0}^*\Gamma^*_{\pi^0}$ are small compared to the squared energy of the 
$\pi^0$ that enters the propagator in (\ref{eq:effprop}). Therefore our results
are not very sensitive to the in-medium changes of the $\pi^0$.\\

\begin{figure} 
\begin{center}
\includegraphics[width=12cm]{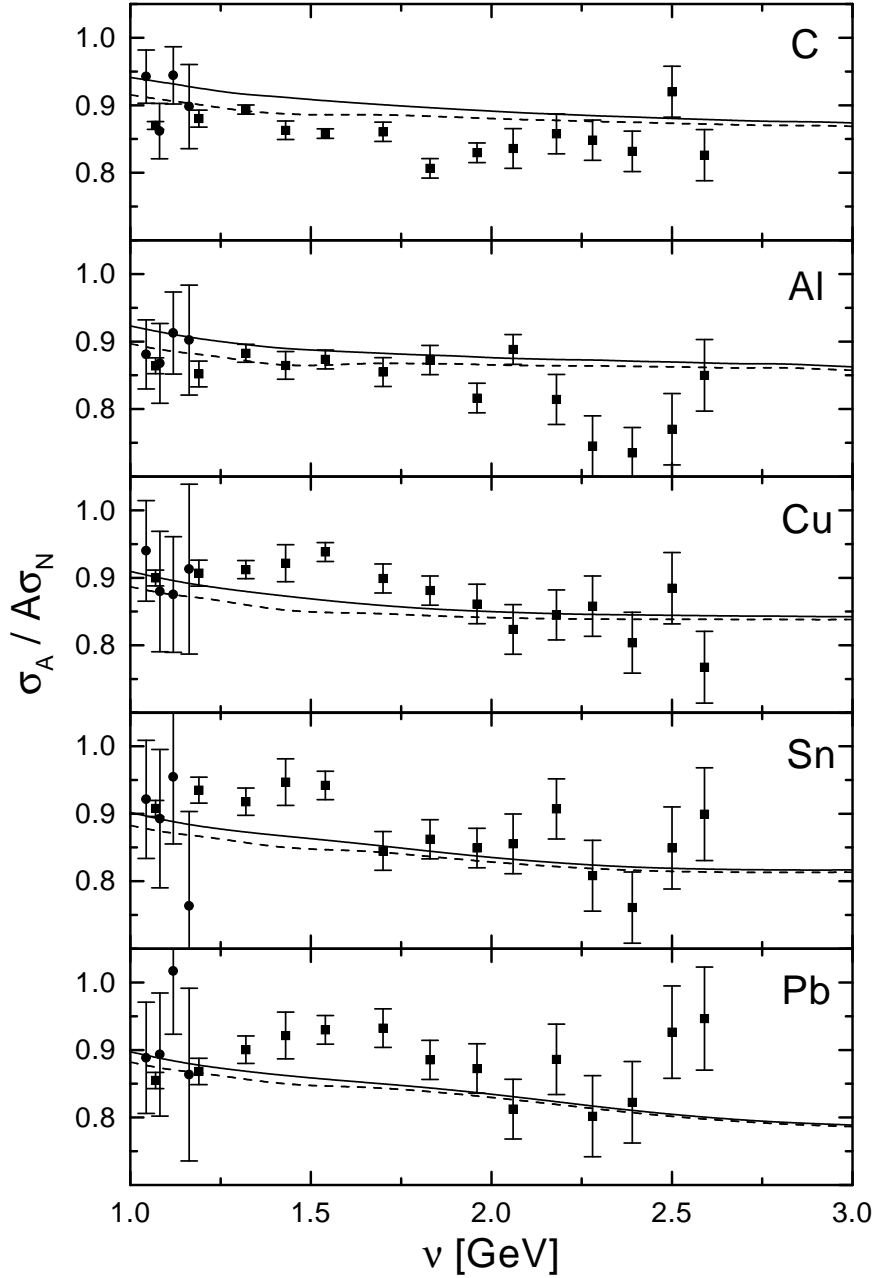}
\end{center}
\caption{Calculated ratio $\sigma_{\gamma A}/A\sigma_{\gamma N}$ as a 
function of the photon energy $\nu$. The solid line corresponds to the 
calculation up to triple scattering involving only the vector mesons. 
The dashed line represents the calculation 
including the $\pi^0$ contribution to nuclear shadowing (\ref{eq:pion}). The 
experimental data are taken from: {\small $\bullet$}~\protect\cite{Bia96}, 
{\tiny $\blacksquare$}~\protect\cite{Muc98}.}
\label{fig:pion}
\end{figure}

Taking the $\pi^0$ contribution into account leads to an enhancement of the
shadowing effect at small energies as can be seen in Fig.~\ref{fig:pion}. The
solid line represents the old result including only $\rho^0$, $\omega$ and
$\phi$ as intermediate states and including terms up to triple scattering in
the multiple scattering series. The dashed line shows the result one gets by 
including the $\pi^0$ as an intermediate hadronic state. At small energies 
and light nuclei the $\pi^0$ contribution (\ref{eq:pion}) to nuclear shadowing
is about 30\%. Here the $\rho^0$ photoproduction, even in forward direction, 
is suppressed by the nuclear formfactor because of its much larger mass. 
Because of angular momentum conservation the lighter $\pi^0$ cannot be produced
in the forward direction without exciting the nucleus. The necessary transverse
momentum transfer increases with rising
photon energy. Therefore the nuclear formfactor suppresses the $\pi^0$ 
contribution to nuclear shadowing at high energies. This
suppression is larger for the heavier nuclei as one would expect from the
$A$ dependence of the nuclear formfactor. This reduces the $\pi^0$ contribution
to shadowing in the case of Pb to about 10\% at small energies.\\


\section{Summary} \label{sec:summary}
In this work we have presented a description of nuclear shadowing in 
photoabsorption at low energies that is more realistic than the simple Glauber
model. We use a multiple scattering expansion to express the nuclear forward 
Compton amplitude in terms of nucleonic scattering amplitudes. From that the 
total nuclear cross section is calculated via the optical theorem. The main 
contribution to nuclear shadowing results from the double scattering term. Here
the incoming photon produces an intermediate hadronic state at one nucleon that
scatters on a second nucleon into the outgoing photon. In contrast to the 
simple Glauber model we also allow for off-shell vector mesons as intermediate
states and avoid the eikonal approximation that is usually made in such 
calculations. We find that very light $\rho^0$ mesons below their pole mass 
$m_\rho=770\text{ MeV}$ give the biggest contribution to nuclear shadowing at 
low energies. This is reasonable because these light states have the largest 
coherence length among the intermediate vector meson states. 
By including only the light vector mesons $\rho$, $\omega$ and $\phi$ as 
intermediate states in the multiple scattering series, we can already reproduce
the photoabsorption data for heavy nuclei. At low energies the shadowing effect
is enhanced by intermediate $\pi^0$. Since a $\pi^0$ has no spin
it cannot be produced in the forward direction without excitation of the 
nucleus. Therefore $\pi^0$ contribution to nuclear shadowing in photoabsorption
is absent in any calculation based on the eikonal approximation. In the present
work we have shown that at low energies intermediate $\pi^0$ are responsible 
for about 10-30\% of the total shadowing effect. At high energies and for heavy
nuclei their contribution is suppressed by the nuclear formfactor.
\\

\acknowledgements
This work was supported by BMBF.


\begin{table}
  \begin{center}
    \begin{tabular}{c||c|c|c|c|c}
      & $^{12}$C & $^{27}$Al & $^{63}$Cu & $^{120}$Sn & $^{208}$Pb\\
      \hline $R/\textrm{fm}$ & $2.209$ & $3.090$ & $4.313$ & $5.513$ & $6.755$\\
      $a/\textrm{fm}$ & $0.479$ & $0.478$ & $0.477$ & $0.476$ & $0.476$\\
      $\rho_0/\textrm{fm}^{-3}$ & $0.182$ & $0.177$ & $0.167$ & $0.159$ & $0.154$
    \end{tabular}
  \end{center}
  \caption[]{Parameters~\protect\cite{Len99} used in the Woods-Saxon distribution~(\ref{eq:woods-saxon}).}
  \label{tab:woods-saxon}
\end{table}


\end{document}